# Non-parallel Voice Conversion System with WaveNet Vocoder and Collapsed Speech Suppression

**YI-CHIAO WU[1] (Student Member, IEEE), PATRICK LUMBAN TOBING[1] (Member, IEEE), KAZUHIRO KOBAYASHI[2] (Member, IEEE), TOMOKI HAYASHI[3], and TOMOKI TODA[2] (Senior Member, IEEE)**

[1]Graduate School of Informatics, Nagoya University, Nagoya 464-8601, Japan
[2]Information Technology Center, Nagoya University, Nagoya 464-8601, Japan
[3]Graduate School of Information Science, Nagoya University, Nagoya 464-8601, Japan

Corresponding author: Yi-Chiao Wu (e-mail: yi-chiao@g.sp.m.is.nagoya-u.ac.jp).

This work was partly supported by JST, CREST Grant Number JPMJCR19A3, JST, PRESTO Grant Number JPMJPR1657, and JSPS KAKENHI Grant Number 17H06101.

**ABSTRACT** In this paper, we integrate a simple non-parallel voice conversion (VC) system with a WaveNet (WN) vocoder and a proposed collapsed speech suppression technique. The effectiveness of WN as a vocoder for generating high-fidelity speech waveforms on the basis of acoustic features has been confirmed in recent works. However, when combining the WN vocoder with a VC system, the distorted acoustic features, acoustic and temporal mismatches, and exposure bias usually lead to significant speech quality degradation, making WN generate some very noisy speech segments called collapsed speech. To tackle the problem, we take conventional-vocoder-generated speech as the reference speech to derive a linear predictive coding distribution constraint (LPCDC) to avoid the collapsed speech problem. Furthermore, to mitigate the negative effects introduced by the LPCDC, we propose a collapsed speech segment detector (CSSD) to ensure that the LPCDC is only applied to the problematic segments to limit the loss of quality to short periods. Objective and subjective evaluations are conducted, and the experimental results confirm the effectiveness of the proposed method, which further improves the speech quality of our previous non-parallel VC system submitted to Voice Conversion Challenge 2018.

## I. INTRODUCTION

Voice conversion is a technique to change speech characteristics such as the speaker identity and emotion of an input speech while maintaining the same linguistic content. In this paper, we focus on general speaker voice conversion to convert a source speaker identity to a specific target speaker. For simplicity, we use VC to refer to speaker voice conversion. Conventional VC models [1–8] are usually trained with a parallel corpus, which consists of source and target training data with the same linguistic contents. However, collecting a huge amount of parallel corpus data for VC training is impractical, so many non-parallel VC methods [9–25] have been proposed. Furthermore, as a more general non-parallel VC application, many cross-lingual VC [26–32] techniques have also been explored.

Conventional VC techniques are usually combined with signal-processing-based (conventional) vocoders such as STRAIGHT [33] and WORLD [34], which encode (analyze) speech into acoustic features such as spectral and prosodic features and decode (synthesize) speech on the basis of these acoustic features. For instance, statistical models such as the Gaussian mixture model (GMM) [1, 2], deep neural network (DNN) [3–5], and exemplar-based models [6–8] have been proposed to convert source acoustic features to target acoustic features, and then the converted waveforms are synthesized on the basis of the converted acoustic features by conventional vocoders. However, the oversimplified assumptions of the speech generation mechanism, such as the fixed length of analysis windows, a time-invariant linear filter, and a stationary Gaussian process, imposed on conventional vocoders lead to loss of phase and temporal details of the original speech, which cause significant speech quality degradation of the synthesized speech signals.

Recently, many autoregressive (AR) techniques directly modeling raw speech waveforms such as WaveNet (WN) [35] and SampleRNN [36] have been proposed, which achieve high-fidelity speech generation by modeling the conditional probability distribution of each speech sample on the basis of past speech samples. In [37, 38], Tamamori and coworkers applied WN to replace the synthesis part of conventional vocoders, which generates speech waveforms on the basis of acoustic features, to markedly improve the quality of the synthesized speech. Specifically, the WN vocoder generates speech conditioned on not only previous speech samples but also conventional-vocoder-extracted acoustic features without various handcrafted assumptions, so the lost phase and temporal details can be greatly recovered by the WN vocoder to improve the quality of the generated speech. Moreover, the results in [37, 38] also demonstrate the effectiveness of the WN vocoder with a small training data set, which greatly reduces the training data requirement of the original WN model.

However, directly combining the WN vocoder with a VC process causes serious mismatch problems. Specifically, because of the length difference between the source and target data of a VC speaker pair, the WN vocoder is usually trained with natural target acoustic feature and waveform pairs. In the testing stage, the trained WN vocoder is conditioned on the converted acoustic features, which have the same data length as the source acoustic features, to generate the converted waveforms, so the acoustic mismatch between the natural and converted acoustic features leads to significant quality degradation such as a waveform-based discontinuity. If we apply a data alignment technique such as dynamic time warping (DTW) to train the WN vocoder with the aligned converted acoustic features and natural waveforms, it will introduce an extra alignment error into the system without solving the temporal mismatch problem [39], which means that although the data are aligned, the phase and temporal details of the converted acoustic features and target waveform signals are still mismatched. Moreover, the inherent exposure bias problem [40, 41], which is caused by the AR nature of the WN vocoder, sometimes leads to unexpected noisy segments, especially when the WN vocoder is conditioned on artificial acoustic features such as those used in VC. In conclusion, the discontinuous waveform signals and unexpected noisy speech segments caused by the acoustic and temporal mismatches and the exposure bias are called the collapsed speech problem [42].

To address this problem, we propose a distribution constraint [42] to directly refine the predicted probability distribution of each speech sample from the output of the WN vocoder, which significantly alleviates the collapsed speech problem. Specifically, the conventional-vocoder-generated speech is a good reference, which is usually stable and collapsed-speech-free, so the predicted distributions of the WN vocoder can be constrained by the sequential correlations of the reference speech. That is, the linear prediction coding (LPC) coefficients of WORLD-generated speech are extracted and the LPC distribution constraint (LPCDC) for each WN-predicted probability distribution is derived from the LPC coefficients and past WN-generated samples. However, the sequential correlations of WORLD-generated waveforms suffer from oversmoothing and quality degradation because of the statistical nature of LPC and the lost phase and details problem of WORLD. These negative effects also degrade the quality of the WN-generated speech when the LPCDC is applied.

To tackle the problem, we propose a collapsed speech segment detector (CSSD) [42] to only apply the LPCDC to the detected collapsed segments, which limits the negative effects of the LPCDC to few speech segments and markedly eases the oversmoothing and quality degradation problems of the LPCDC. In this paper, the proposed LPCDC w/ CSSD approach is evaluated with a baseline non-parallel VC system [43], which was submitted to the intralanguage non-parallel VC task (SPOKE) of Voice Conversion Challenge 2018 (VCC2018) [44].

Compared with our previous works [42, 43], the new contributions of this paper are as follows:

- We conduct a comprehensive subjective evaluation to further explore the degree of speech quality degradation caused by WORLD, collapsed speech, the LPCDC, and the proposed LPCDC w/ CSSD, while the previous works only showed the effectiveness of the proposed LPCDC w/ CSSD.
- We explore the probability distribution of collapsed speech and provide an analytical explanation of the reason for the collapsed speech problem and the effectiveness of our proposed method.
- We improve the speech quality of our previous non-parallel VC system submitted to VCC2018.

The paper is organized as follows. In Section II, we review the related works. In Section III, a brief introduction to the baseline non-parallel VC system with the WN vocoder is presented. In Section IV, we describe the concepts and details of the proposed LPCDC and CSSD. In Section V, we report objective and subjective tests carried out to evaluate the effectiveness of the proposed LPCDC w/ CSSD. Finally, the conclusion is given in Section VI.

## II. RELATED WORKS

### A. INTRALANGUAGE NON-PARALLEL VC
As a practical application, non-parallel VC has been explored by many different approaches. For example, Erro et al. [9] proposed the INCA algorithm to iteratively align the non-parallel corpus for conventional parallel GMM-based VC. Sun et al. [10] and Xie et al. [11] proposed similar frameworks using a well-trained automatic speech recognition (ASR) model to extract speaker-independent (SI) phonetic posteriorgrams (PPGs) and adopt a speaker-dependent (SD) PPG-to-spectrum model to generate converted spectra. Restricted Boltzmann machine (RBM)- [12] and variational autoencoder (VAE)-based [13–15] models have also been proposed to disentangle the acoustic features into SD and SI components for VC. Moreover, inspired by the success of cycle consistent adversarial networks (CycleGAN) for image translation [45], cycle-consistency has been widely applied to non-parallel VC [16–18].

In addition, non-parallel VC with external reference speakers has also been widely surveyed. For instance, building a VC model of reference speakers with a parallel corpus and adapting it for source and target speakers with a non-parallel corpus [19, 20] achieved early success. Inspired by the GMM-based speaker verification technique [46], speaker adaptation from a universal background model (UBM) [21] for non-parallel VC also shows the effectiveness of assistance from reference speakers. Representing the source and target utterances on the basis of weighted reference dictionaries also attains good quality for exemplar-based VC [22]. In this paper, we utilize a two-stage non-parallel VC with a reference speaker that is based on a system with a combination of many-to-reference and reference-to-many models [23–25].

### B. NEURAL VOCODER
Recently, AR models, which adopt a recurrent neural network (RNN) [36] or deep dilated convolutional neural network (CNN) [35] to directly model raw speech waveform signals with a very high temporal resolution, have achieved significantly high quality speech generation performance. Specifically, the AR models predict the probability of the current speech sample on the basis of previous samples and auxiliary features such as acoustic features. However, because of the modeling of very long term correlations of speech samples and the AR nature, the huge network and computation time requirements make AR models impractical for real-world applications. Therefore, novel network architectures with speech and signal processing domain knowledge such as FFTNet [47], WaveRNN [48], QPNet [49], and glottal-excitation-related [50, 51] and LPC-related [52] architectures have been proposed to reduce the requirements for speaker adaptation and network capacity. Furthermore, flow-based models adopting inverse AR flow [53, 54] and Glow [55, 56] architectures also markedly reduce the generation time with non-AR approaches. Non-AR vocoders with mixed-excitation-like signals [57, 58] also achieve very high speech quality with a much higher synthesis speed than the AR models. However, in this paper, we focus on collapsed speech detection and suppression, so we apply the proposed method with the classical WN model.

### C. VC WITH WAVENET VOCODER
Because both spectral and prosodic features are speaker-dependent, manipulating the speaker identity in the acoustic feature space then synthesizing the converted speech with the modified acoustic features via a vocoder is a conventional VC flow. However, the oversimplified assumptions of the speech generation mechanism of conventional vocoders usually cause severe speech quality degradation. To solve this problem, the WN vocoder [37, 38], which markedly improves the speech quality of the synthesized waveforms, has been proposed. Kobayashi et al. [59] applied the WN vocoder to a GMM-based VC system and achieved higher speech quality and similarity than the WORLD vocoder. Furthermore, we also applied the WN vocoder to DNN- [43], deep mixture density network (DMDN)- [60], long short-term memory (LSTM)- [61], VAE- [62], and gated recurrent unit (GRU)-based [18] VC models, which also confirmed the effectiveness of the WN vocoder. Liu et al. [63] and Sisman et al. [64] applied the WN vocoder to their conversion models with the assistance of PPGs and also attained significant improvements compared with conventional vocoders.

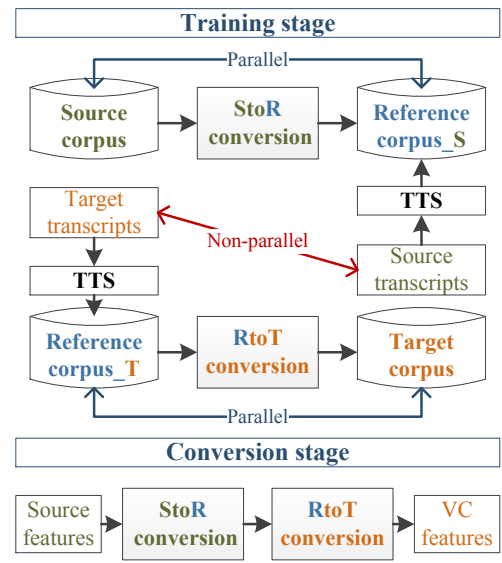

**FIGURE 1.** Two-stage non-parallel spectral feature conversion.

## III. BASELINE NON-PARALLEL VC SYSTEM
In this section, we first introduce the baseline two-stage non-parallel VC with a reference speaker, which was applied to our previous system submitted to the SPOKE task of VCC2018. Then, we present the DNN- and DMDN-based [65] VC models. Finally, we introduce the WN vocoder.

### A. NON-PARALLEL VC WITH REFERENCE SPEAKER
Parallel VC models are usually trained with a parallel corpus, which has an inherent one-to-one relationship between source and target data, to construct a source-to-target mapping function. However, an inherent one-to-one relationship does not exist in a non-parallel corpus such as the SPOKE set of VCC2018. To address this issue, since the transcripts of the SPOKE set are available, it is feasible to use a TTS system to generate corresponding parallel utterances for the source and target speakers of the SPOKE set. That is, we take the TTS speaker as the reference speaker to develop the source-to-reference (StoR) and reference-to-target (RtoT) models, which are parallel VC models, in the training stage and convert the source features to target features via the cascaded StoR and RtoT models. As shown in Fig. 1, the parallel corpus for training the StoR model includes the source corpus and reference corpus_S, which is established by a unit-selection-based single-speaker TTS system with the source transcripts. The RtoT model is trained with the parallel corpus including the target corpus and reference corpus_T, which is established by the same TTS system but with the target transcripts. In the conversion stage, the input source features are converted to reference features by the StoR model, and then the reference features are further converted to the target features by the RtoT model. Finally, the WN vocoder generates the converted speech on the basis of the converted acoustic features.

Furthermore, to alleviate the alignment mismatch between the source/target and reference utterances, the human-labeled short pauses and silences of the training utterances are adopted to handle the short pauses and silences of the TTS-generated speech to match it with the corresponding source/target speech. After that, because the TTS system uses hidden Markov model (HMM)-state alignments, the framewise DTW technique is still applied to alleviate the spectral mismatch between the natural acoustic features and the acoustic features extracted from the TTS-generated speech. In conclusion, arbitrary parallel VC models can be adopted for non-parallel VC using the TTS-generated parallel corpus and cascaded two-stage approach.

The VC models introduced in this paper focus on spectral conversion, and the pitch is linearly transformed.

### B. DNN-BASED VC

In the non-parallel VC system [43] submitted to VCC2018, a DNN-based framewise spectral conversion model is adopted. Specifically, given the source feature vector $\mathbf{S}_n = \left[ \mathbf{s}_n^\mathrm{T}, \Delta \mathbf{s}_n^\mathrm{T} \right]^\mathrm{T}$ and target feature vector $\mathbf{T}_n = \left[ \mathbf{t}_n^\mathrm{T}, \Delta \mathbf{t}_n^\mathrm{T} \right]^\mathrm{T}$, which include static and delta spectral features with the frame index $n$, the DNN models the conditional probability formulated as

$$P\left( \mathbf{T}_n \mid \mathbf{S}_n, \boldsymbol{\Sigma}, \boldsymbol{\lambda} \right) = \mathrm{N}\left( \mathbf{T}_n ; \mathrm{f}_\lambda \left( \mathbf{S}_n \right), \boldsymbol{\Sigma} \right), \quad (1)$$

where $\boldsymbol{\lambda}$ and $\mathrm{f}_\lambda(\cdot)$ respectively denote the parameters and nonlinear transformation function of the DNN model, $\mathrm{N}(\cdot)$ is the Gaussian distribution, and $\boldsymbol{\Sigma}$ is the diagonal covariance matrix of the training data. In the training stage, the DNN parameter $\hat{\boldsymbol{\lambda}}$ is estimated as

$$\hat{\boldsymbol{\lambda}} = \underset{\boldsymbol{\lambda}}{\operatorname{argmax}} \sum_{n=1}^{N} \log P\left( \mathbf{T}_n \mid \mathbf{S}_n, \boldsymbol{\Sigma}, \boldsymbol{\lambda} \right)$$
$$= \underset{\boldsymbol{\lambda}}{\operatorname{argmin}} \frac{1}{2} \sum_{n=1}^{N} \left( \mathbf{T}_n - \mathrm{f}_\lambda \left( \mathbf{S}_n \right) \right)^\mathrm{T} \boldsymbol{\Sigma}^{-1} \left( \mathbf{T}_n - \mathrm{f}_\lambda \left( \mathbf{S}_n \right) \right). \quad (2)$$

In the conversion stage, to alleviate the discontinuity caused by the framewise approach and the oversmoothing effect caused by the statistical nature, the maximum likelihood parameter generation (MLPG) [66] and global variance (GV) postfilter [2] techniques are adopted.

### C. DMDN-BASED VC

In this paper, a DMDN model [65] is applied to our two-stage non-parallel VC system, which models the conditional probability with mixtures of Gaussian distributions instead of a single Gaussian distribution. The multimodal approach with the variance predicting capability enhances the model capacity. Given the same condition as (1), the DMDN-based conditional probability is formulated as

$$P\left( \mathbf{T}_n \mid \mathbf{S}_n, \theta \right) = \sum_{m=1}^{M} \alpha_m \left( \mathbf{S}_n \right) \mathrm{N}\left( \mathbf{T}_n \mid \mu_m \left( \mathbf{S}_n \right), \sigma_m^2 \left( \mathbf{S}_n \right) \right), \quad (3)$$

where $\theta$ denotes the parameters of the DMDN model, $\mathrm{N}\left( \cdot \mid \mu, \sigma^2 \right)$ denotes a single Gaussian mixture with the mean $\mu$ and covariance matrix $\sigma^2$, $M$ is the total number of mixture components, $m$ is the mixture index, and $\alpha_m \left( \mathbf{S}_n \right)$ denotes the mixture weight of the $m$th component given $\mathbf{S}_n$. As a result, the outputs of the network are as follows:

$$\alpha_m \left( \mathbf{S}_n \right) = \frac{\exp\left( z_m^{(\alpha)} \left( \mathbf{S}_n, \theta \right) \right)}{\sum_{j=1}^{M} \exp\left( z_j^{(\alpha)} \left( \mathbf{S}_n, \theta \right) \right)}, \quad (4)$$

$$\mu_m \left( \mathbf{S}_n \right) = z_m^{(\mu)} \left( \mathbf{S}_n, \theta \right), \quad (5)$$

$$\sigma_m \left( \mathbf{S}_n \right) = \exp\left( z_m^{(\sigma)} \left( \mathbf{S}_n, \theta \right) \right), \quad (6)$$

where $z^{(\alpha)}$, $z^{(\mu)}$, and $z^{(\sigma)}$ respectively denote the weight, mean, and variance activations of the DMDN output layer. The updated form of the DMDN parameters is defined as

$$\hat{\theta} = \underset{\theta}{\operatorname{argmax}} \sum_{n=1}^{N} \log P\left( \mathbf{T}_n \mid \mathbf{S}_n, \theta \right), \quad (7)$$

Moreover, the MLPG and GV techniques are also adopted in the DMDN conversion stage.

### D. WAVENET VOCODER

To model very high temporal resolution speech waveform signals, WN [35] adopts an AR approach, which generates the speech waveform sample by sample. Specifically, WN models the probability distribution of each speech sample conditioned on a segment of previous samples called a *receptive field*. To guide WN to generate the desired speech content, WN is conditioned on not only the *receptive field* but also auxiliary features such as linguistic features. Furthermore, taking WN as a vocoder [37, 38], which adopts the conventional-vocoder-extracted acoustic features as the auxiliary features, greatly reduces the huge training data requirement and makes it easy to combine WN with conventional VC systems [59–65].

The conditional probability of the WN vocoder is formulated as

$$P\left( \mathbf{Y} \mid \mathbf{h} \right) = \prod_{t=1}^{T} P\left( y_t \mid y_{t-1}, \ldots, y_{t-r}, \mathbf{h} \right), \quad (8)$$

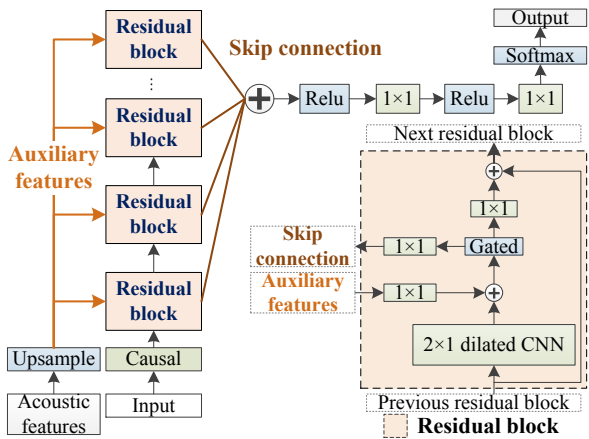

**FIGURE 2.** WaveNet vocoder architecture.

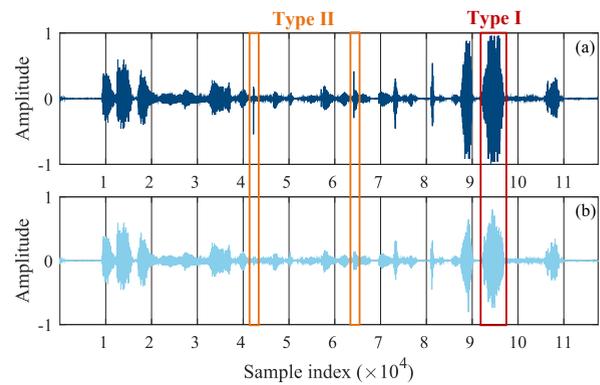

**FIGURE 3.** (a) WN-generated waveform w/ collapsed speech. (b) WN-generated waveform w/ LPCDC and CSSD.

where $t$ is the sample index, $r$ is the length of the *receptive field*, $y_t$ is the current audio sample, and **h** is the vector of the acoustic features. To model the long-term dependence among speech samples, WN applies a deep architecture including many residual blocks as shown in Fig. 2. Specifically, each residual block contains a gated structure with a skip connection, an auxiliary feature condition, and a dilated CNN layer. The dilation size of each residual block is different, which makes the network capture information on different levels and efficiently extend the *receptive field*. The stacked residual blocks with skip connections capture the hierarchical information of the speech samples in the *receptive* field and send them to the output layer to predict the probability distribution of the current sample. In this paper, the μ–law is applied to encode speech waveforms into 8 bits, so the output of the WN vocoder is a logistic distribution with 256 levels.

## IV. WAVENET VOCODER WITH COLLAPSED SPEECH SUPPRESSION AND DETECTION

In this section, we first present the collapsed speech problem, which has two types of collapsed waveform. Then, we describe the proposed LPCDC and CSSD for the suppression and detection of collapsed speech.

### A. COLLAPSED SPEECH PROBLEM

Although the effectiveness of the WN vocoder for generating high-fidelity speech on the basis of acoustic features has been proved, the AR nature and waveform-based modeling make the WN vocoder vulnerable to prediction errors. Specifically, because the WN vocoder is conditioned on previous samples to predict the current sample, a prediction error will propagate through the sequential speech samples. The negative ripple effect easily leads to the WN vocoder generating very noisy speech, which is similar to white noise, especially when conditioned on acoustic features with high amplitudes. This white-noise-like speech is defined as Type I collapsed speech as shown in Fig. 3 (a). Furthermore, even if the prediction error problem only occurs in a few samples because of the guide from the acoustic features, it still leads to the WN vocoder generating short impulse noise, which causes significant perceptual quality loss. We define the short impulse noise as Type II collapsed speech as shown in Fig. 3 (a).

The possible reasons for collapsed speech are a lack of training data, conditioning on artificial acoustic features, and exposure bias [40], [41]. Specifically, because of the limited training data or the acoustic mismatch between the training and testing data (e.g., training the WN vocoder with natural acoustic features but testing it with artificial ones), the testing data of the WN vocoder are unseen data, which usually leads to the WN vocoder generating unexpected speech waveforms. Furthermore, even if we apply a data alignment technique such as DTW to avoid the data length mismatch between the source and target data for VC, which allows the WN vocoder to be trained with the data pair of VC acoustic features and natural waveforms, it will introduce an extra alignment error and will not remove feature-waveform temporal mismatch [39] problems such as phase mismatch. Moreover, because the AR WN model is usually trained with ground-truth natural waveforms but tested with self-generated waveforms, the different decoding behavior, which is called the exposure bias problem, sometimes leads to unexpected generation results.

### B. LPC DISTRIBUTION CONSTRAINT

To tackle the collapsed speech problem, we propose the LPCDC to constrain the output probability distribution of the WN vocoder, which is a postprocessing module to prevent the WN vocoder from generating unexpected noise segments. Specifically, according to our observations, although the naturalness of WORLD-generated speech is lower than that of WN-generated speech, it seldom suffers from the collapsed speech problem and has a higher continuity. Moreover, WORLD-generated speech is available when acoustic features, which are the auxiliary features for the WN vocoder, exist. Therefore, the proposed LPCDC extracts the correlations, which are described via the LPC coefficients, of the WORLD-generated speech samples and constrains the corresponding outputs of the WN vocoder with these correlations.

As shown in Fig. 4, the LPCDC-constrained (modified) form of (8) is derived as

$$P(y_t \mid y_{t-r}, \ldots, y_{t-1}, \mathbf{h}, \phi) \propto$$

$$P(y_t \mid y_{t-r}, \ldots, y_{t-1}, \mathbf{h})\big(P(y_t \mid y_{t-l}, \ldots, y_{t-1}, \phi)\big)^{\rho}, \quad (9)$$

where $l$ is the number of LPC dimensions, $\phi$ denotes the LPC coefficients, which are extracted from the corresponding WORLD-generated speech, and $\rho$ is a regularization hyperparameter. That is, the probability distribution of each speech sample is constrained by the LPCDC mask $\big(P(y_t \mid y_{t-l}, \ldots, y_{t-1}, \phi)\big)^{\rho}$, which is a probability mass function (PMF) approximating a Gaussian distribution with the mean $\mu_{LPC}$ and variance $\sigma_{LPC}^2$. The mean $\mu_{LPC}$ is the LPC-predicted value of the current sample, which is given by the weighted sum of the past samples multiplied by the $l$-dimensional LPC coefficients. The variance $\sigma_{LPC}^2$ is the variance of the prediction errors derived from the corresponding frame of the WORLD-generated speech utterance. In addition, $\rho$ is the weight used to control the balance between the LPCDC mask and the WN-predicted probability distribution. More details of the LPC coefficient extraction and LPCDC mask derivation can be found in Appendix.

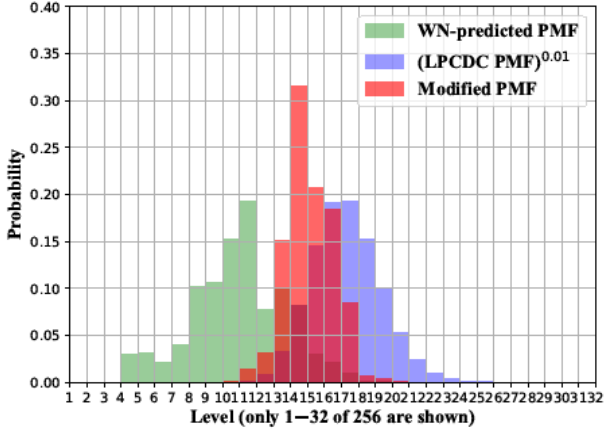

**FIGURE 4.** Probability distributions of WN-predicted PMF, LPCDC PMF with regularization ρ=0.01, and LPCDC-modified PMF.

### C. COLLAPSED SPEECH SEGMENT DETECTION

Although the proposed LPCDC markedly alleviates the collapsed speech problem, the phase and temporal detail loss of the WORLD vocoder and the statistical nature of the LPC lead to speech quality degradation and oversmoothing when deriving the LPCDC mask from the WORLD-generated speech. Therefore, we propose the CSSD to segmentally detect the collapsed speech and only apply the LPCDC to the detected segments that suffer from the collapsed speech problem. Because an utterance usually contains only a few collapsed segments, this mechanism restricts the negative effects of the LPCDC to very short periods. The CSSD also improves the generation efficiency because only the detected segments are regenerated with LPCDC instead of the whole utterance.

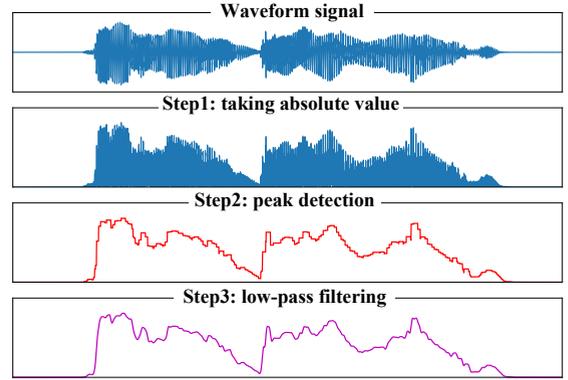

**FIGURE 5.** Three steps of waveform shape detection.

The core role of the CSSD is to segmentally compare the WN-generated waveform envelope with the WORLD-generated reference waveform envelope, and a segment is detected as a collapsed segment when the difference between the two envelopes is larger than a threshold, which is determined by a detection error tradeoff (DET) curve. Note that because the conditional acoustic features of the WN vocoder already contain a power component, which is consistent with that of the acoustic features for WORLD synthesis, the amplitudes of the WN- and WORLD-generated waveform envelopes should be similar. Therefore, the CSSD is employed without any waveform normalization.

Moreover, because of the frequent detection requirements, we adopt a low-computational-cost approach [67] to obtain the waveform envelopes for the CSSD. As shown in Fig. 5, we first take the absolute value of waveform signals. Secondly, a peak detection is performed by dividing the whole absolute sequence into non-overlapping slots and replacing all signals in each slot with the one with the maximum value in that slot. Finally, the final waveform envelope is obtained by processing the detected peak sequence with a low-pass filter. Furthermore, in this paper, we adopt the Hilbert transform (HT) instead of taking the absolute value in the first step because of the lower collapsed speech detection error, which will be demonstrated in Section V–B. More details of the hyperparameters of the CSSD can be found in Section V–A.

### D. WN VOCODER WITH LPCDC AND CSSD

The proposed system, which includes the WN and WORLD vocoders and the LPCDC and CSSD modules, is shown in Fig. 6. Given a sequence of acoustic features, the WORLD vocoder first generates the reference utterance for the LPC coefficients and reference envelope extractions. Then, the CSSD checks every WN-generated speech segment while the generation of this segment is completed by the WN vocoder. The proposed system will automatically regenerate the speech waveforms with an increased regularization $\rho$ up to three times while the collapsed speech problem is detected, and the system preserves the latest results. $\rho$ is set as 0.01, 0.1, and 1 for the first, second, and third regenerations, respectively. The additional computational costs of the proposed system are mainly from WN regenerations compared with other fast modules such as LPC extraction and LPCDC distribution derivation, waveform envelope detections and comparisons for CSSD, and WORLD synthesis. Therefore, if the WN generation time can be markedly reduced, a robust low-latency segmental generation system might be implemented on the basis of the proposed system.

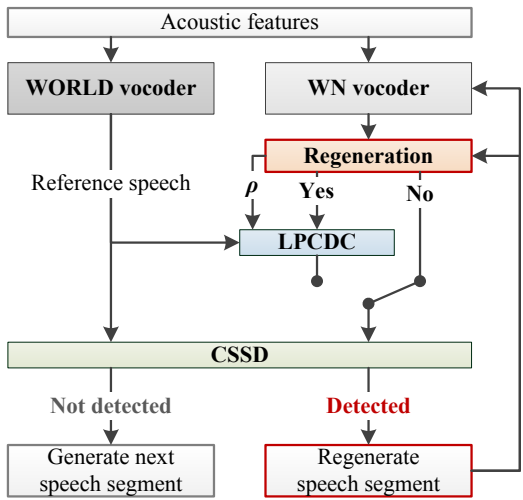

**FIGURE 6.** Proposed WN vocoder with LPCDC and CSSD.

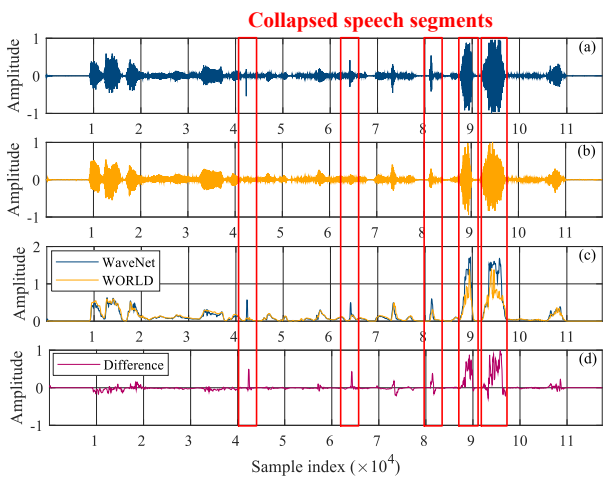

**FIGURE 7.** (a) WN-generated waveform w/ collapsed speech. (b) WORLD-generated waveform. (c) Extracted waveform envelopes. (d) Difference in waveform envelope.

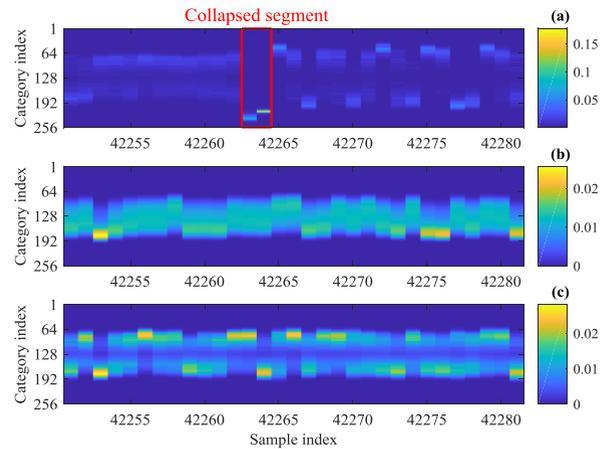

**FIGURE 8.** (a) Predicted PMF sequence of WN w/ Type II collapsed speech. (b) PMF sequence from LPCDC. (c) Modified PMF sequence of WN w/ LPCDC and CSSD.

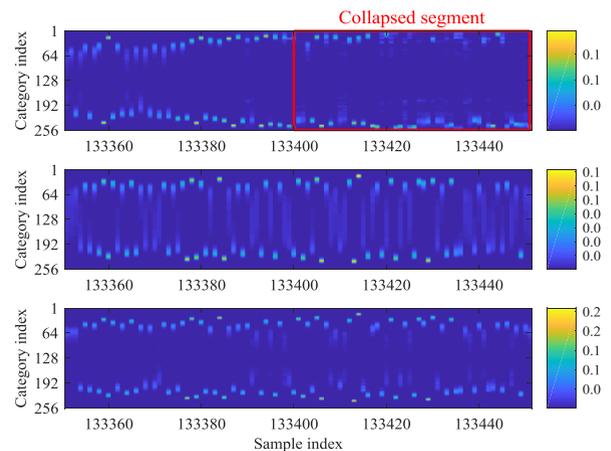

**FIGURE 9.** (a) Predicted PMF sequence of WN w/ Type I collapsed speech. (b) PMF sequence from LPCDC. (c) Modified PMF sequence of WN w/ LPCDC and CSSD.

We take the utterance of Fig. 7 (a) as an example, which is the same utterance as that in Fig. 3 (a). The corresponding WORLD-generated waveform is shown in Fig. 7 (b), the extracted waveform envelopes are shown in Fig. 7 (c), and the difference in waveform envelope is shown in Fig. 7 (d). The WORLD-generated waveform is stable and without collapsed speech segments, but the WN-generated one contains several Type I and II collapsed speech segments. The results in Figs. 7 (c) and (d) confirm the effectiveness of the CSSD module, which detects the collapsed speech segment on the basis of the envelope difference. Moreover, for the PMF sequence shown in Fig. 8 (a), which is part of the first Type II collapsed speech segment in Fig. 7 (a), a few samples with unexpected prediction errors lead to serious unexpected impulse noise. However, after applying the LPCDC PMF sequence shown in Fig. 8 (b) to constrain the WN vocoder outputs, the modified PMF sequence shown in Fig. 8 (c) is free from the unexpected prediction error. The PMF sequences in Fig. 9, which correspond to the last Type I collapsed speech segment in Fig. 7 (a), also show the effectiveness of the proposed system. Specifically, most PMF values of the collapsed segment in Fig. 9 (a) are close to extremums, which represent continuous maximum amplitudes, but the modified PMF sequence in Fig. 9 (c) is more speech-like. Note that the modified predicted PMF sequence of Fig. 8 (c) / 9 (c) is not the result of directly multiplying the predicted PMF sequence of Fig. 8 (a) / 9 (a) by the LPCDC PMF sequence of Fig. 8 (b) / 9 (b). Because of the AR manner of WN, when the first sample in this segment is changed by the LPCDC, the distributions of the following samples are also affected. Finally, the refined speech waveform is shown in Fig. 3 (b), which is free from collapsed speech.

## V. EXPERIMENTAL EVALUATIONS

In this section, we present collapsed speech detection, spectral conversion (objective), and perceptual quality (subjective) evaluations to respectively confirm the effectiveness of the proposed CSSD module, the baseline non-parallel VC model, and the proposed WN vocoder with the LPCDC and CSSD.

### A. EXPERIMENTAL SETTINGS

#### 1) CORPUS

For the experiments, we adopted the VCC2018 [44] corpus, parts of the CMU-ARCTIC [68] corpus, and an internal corpus. The language of these three corpora was English. The VCC2018 corpus included the HUB and SPOKE subsets. The HUB set contained four source speakers and four target speakers with parallel utterances and transcripts, and each speaker had 80 utterances for training and 35 utterances for testing. The SPOKE set also contained four speakers with parallel utterances and transcripts, and each speaker also had 80 utterances for training and 35 utterances for testing. However, the speech contexts of the HUB and SPOKE sets were different, so the VCC2018 SPOKE task, which was a non-parallel VC task, was established with the source speakers from the SPOKE set and the target speakers from the HUB set. Both sets included a balanced number of female and male speakers. All VCC2018 data were recorded in a quiet environment with a sampling rate of 22,050 Hz and 16-bit quantization. In the CMU-ARCTIC corpus, because only speakers "bdl" and "slt" contained 32 kHz data, which was higher than the target sampling rate of 22,050 Hz, only these two speakers' data were involved in the WN vocoder training. Speaker "bdl" had 1131 utterances and speaker "slt" had 1132 utterances. All data of speakers "bdl" and "slt" were downsampled from 32 kHz to 22,050 Hz, and the quantization number was also 16 bits. Furthermore, an internal single male speaker corpus was adopted to build the reference TTS system, which included 3000 utterances.

#### 2) ACOUSTIC FEATURES

The WORLD [34] vocoder was adopted to extract a 513-dimensional aperiodicity ($ap$), 513-dimensional spectral envelope ($sp$), and one-dimensional fundamental frequency ($F_0$) with 25 ms frame length and 5 ms frameshift. The $sp$ feature was further parameterized into a 34-dimensional Mel-spectrum ($mcep$), and the $ap$ feature was coded into a two-dimensional aperiodic component. Joint spectral features were aligned via DTW. For non-parallel VC, each source $mcep$ was converted to a target $mcep$ by the two-stage VC model. A source $F_0$ sequence was linearly transformed into a target one in the logarithm domain. The source $ap$ was kept the same. The auxiliary features of the WN vocoder included $mcep$, coded $ap$, interpolated continuous $F_0$, and a voice/unvoice binary code. The LPC coefficients for the LPCDC were 30-dimensional with 20 ms frame length and 5 ms frameshift.

#### 3) WN VOCODER

The number of residual blocks of the WN vocoder was 30, and all dilated and 1×1 convolutions in the residual blocks had 512

channels. The 1×1 convolutions between the skip connection and softmax had 256 channels. The dilation size was set to $2^0$–$2^9$ with three cycles (one cycle included 10 residual blocks). The number of trainable parameters was 44 million. A multi speaker WN vocoder was trained on the basis of the training data of all VCC2018 speakers and speakers "bdl" and "slt" of the CMU-ARCTIC corpus, and then four SD WN vocoders were fine-tuned by updating the output layers of the multi speaker WN vocoder with the training data of the corresponding target speakers. The number of training iterations was 200,000 and the training learning rate was initially 0.001 with 50% decay per 50,000 iterations. The number of updating iterations was 50,000 and the updating learning rate was 0.001 without decay. The mini-batch size was 20,000 samples and Adam [69] was adopted for optimization. Furthermore, the noise shaping (NS) technique [70] was applied to the WN vocoders.

#### 4) VC MODELS

Both DNN- and DMDN-based VC models contained four hidden layers with 1024 hidden units, and the mixture number of the DMDN model was 16. The learning rates of these two models were $6 \times 10^{-4}$ without decay, the training epoch was 15, the utterance-based mini-batch was adopted, and Adam was also used for optimization.

#### 5) CSSD

The length of the speech segment of the CSSD was 4000 samples, which means that the system checked for collapsed speech every time the WN vocoder generated 4000 new samples. The length of the peak detection window was 200 samples and the cutoff frequency of the low-pass filter in the CSSD was 300 Hz.

### B. COLLAPSED SPEECH DETECTION EVALUATION

To evaluate the performance of the proposed CSSD, a human-labeled test set of the SPOKE task, which was established using DNN-based non-parallel VC models with the WN vocoder, was adopted. The number of speaker pairs of the SPOKE task was 16, so the total number of utterances in this test set was 560. According to the labeled results, 46 utterances suffered from the Type I collapsed speech problem and 276 utterances had the Type II short impulse noise. Although more than 50% of the utterances suffered from the collapsed speech problem, some utterances with the Type II short impulse noise did not cause perceptual degradation. This is because the label criterion was only based on the waveform shape and the perceptual loss was not considered.

Collapsed speech detection is a verification problem, which is measured via the false acceptance rate (FAR) and false rejection rate (FRR). Specifically, we measured the detection performance using the FAR of the collapsed utterances that were not detected and the FRR of the normal utterances that were detected as shown in Fig. 10. We compared the proposed CSSD with Mel-cepstrum distortion (MCD)- and power-based detections, which detected the collapsed speech segments on the basis of MCD and power differences between the generated and reference utterances [43], respectively. Furthermore, because we adopted the NS technique for the WN vocoder and the HT for waveform envelope extraction, the four CSSD variants of the waveform envelope extraction before (bNSR) and after NS restoration (aNSR) and with (w/ HT) and without the HT (w/o HT) were also considered in the comparison.

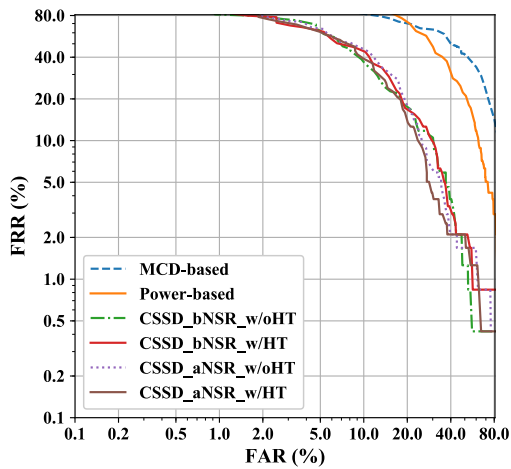

**FIGURE 10.** DET curve for overall collapsed speech detection.

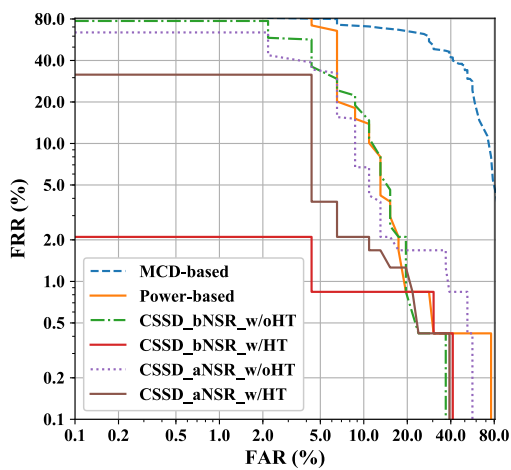

**FIGURE 11.** DET curve for Type I collapsed speech detection.

As shown in Fig. 10, the overall detection performance, which includes the Type I and II collapsed speech segments, indicates that the proposed CSSD significantly outperforms the MCD- and power-based methods. Furthermore, if we only compare the utterances suffering from Type I collapsed speech segments with the normal utterances, as shown in Fig. 11, the CSSD-series methods still achieve a lower equal error rate (EER), especially the methods with the HT. To summarize, the experimental results confirm the effectiveness of the proposed CSSD with the HT, which detects Type I collapsed speech segments with an EER lower than 5% and both Type I and Type II collapsed speech segments with an EER of 20%.

| | DNN-based | | DMDN-based | |
|---|---|---|---|---|
| | w/o GV | w/ GV | w/o GV | w/ GV |
| One-stage | 5.48 | 6.12 | 5.39 | 6.00 |
| Two-stage | 5.64 | 6.22 | 5.59 | 6.13 |
| Non-parallel | 5.54 | 6.01 | 5.46 | 5.90 |

Because of the convenience of implementation and the similar detection performance, the following tests were conducted on the system with the CSSD applied with the HT before the NSR.

### C. OBJECTIVE SPECTRAL MAPPING EVALUATION

To easily evaluate the spectral conversion capability of the proposed two-stage non-parallel VC, we conducted a spectral mapping evaluation with a parallel corpus. Specifically, we took the four speakers of the SPOKE set to form 12 speaker pairs with a parallel corpus, so the parallel VC (one-stage) models, which directly converted the source spectral features to the target ones, were available. Furthermore, the four SPOKE speakers used for reference and four reverse models were trained with the SPOKE set and corresponding TTS-generated utterances. These eight VC models formed 12 simulated non-parallel VC (two-stage) paired models, which did not adopt any source-target parallel information.

As shown in Table I, the DMDN-based VC models achieve a slightly higher spectral prediction accuracy with lower MCD values than the DNN-based VC models. Although applying the GV postfilter leads to higher MCD values, the tendency is still the same. The results confirm that the DMDN-based baseline system attains a reasonable spectral prediction accuracy, which is comparable to that of the previous DNN-based baseline system. Furthermore, although the parallel VC (one-stage) models exhibit a higher conversion performance, the two-stage models still achieve an acceptable conversion accuracy.

In addition, the results for the real non-parallel VC data (non-parallel) utilized in the SPOKE task of VCC2018, are also presented in Table I. The tendency is similar to the simulated results of the parallel data. To summarize, the TTS-generated speech already contains sufficient speech components to be used as the reference speech, which was also confirmed in our previous work [43].

### D. SUBJECTIVE EVALUATION

To evaluate the perceptual performance of the proposed system, we conducted a speech quality evaluation measured by a mean opinion score (MOS) and a speaker similarity evaluation measured by a similarity score. In this subsection, both DNN- and DMDN-based non-parallel VC models were trained with a non-parallel corpus, which took the speakers of the SPOKE set as the sources and the four target speakers of the HUB set as the targets to form 16 speaker pairs. The demo utterances can be found in [71].

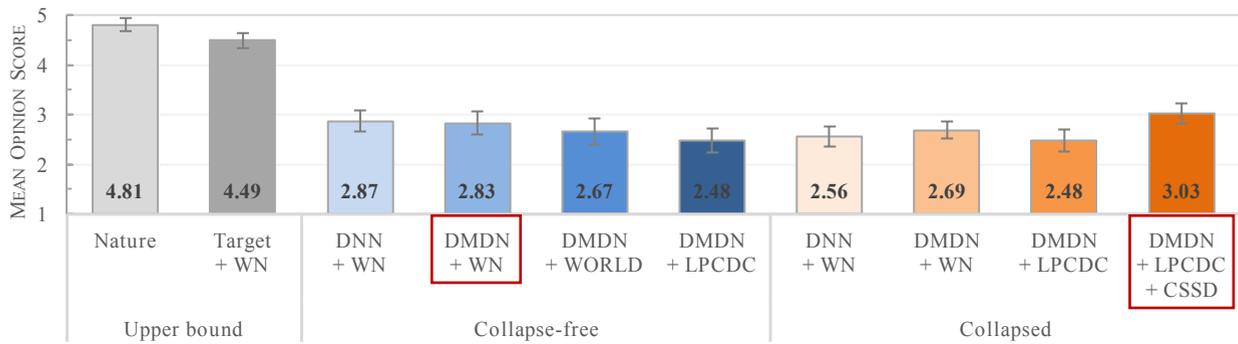

**FIGURE 12.** MOS evaluation of speech quality with 95% confidence intervals. (The performance of the proposed system is the combination of DMDN + WN in the collapse-free set and DMDN + LPCDC + CSSD in the collapsed set.)

### 1) SPEECH QUALITY

In the speech quality test, we compared 10 systems from three subsets, which were upper bound, collapse-free, and collapsed sets. The upper bound set included natural speech and the WN vocoder conditioned on the target natural acoustic features. The collapse-free set included DNN- and DMDN-based non-parallel VC models with different vocoders and conditions, and for all the generated utterances in this set, no collapsed speech segments were detected by the CSSD. The collapsed set also included DNN- and DMDN-based non-parallel VC models combined with the WN vocoder and several conditions, but all the utterances contained detected the collapsed speech segments. Collapsed utterances were detected in 377 of 560 utterances generated by the DNN-based system, and the number of detected collapsed utterances of the DMDN-based system was 335. Possible reasons for the high ratio of collapsed utterances were the 20% EER of the CSSD and the unoptimized threshold.

We randomly selected five utterances of each system and speaker pair to form an evaluation set, which included 800 (5 × 16 × 10) utterances. Then, we divided the evaluation set into five subsets, each set containing the same number of utterances under each condition. Moreover, each subset was evaluated by three listeners with the same device in a quiet environment, so 15 listeners took part in this evaluation. Although the listeners were not native English speakers, they had worked on speech or audio generation research. The speech quality of each utterance was evaluated by the listeners, who assigned an MOS of 1–5, where the higher the MOS, the higher the speech quality of the utterance.

As shown in Fig. 12, although the synthesized speech of the WN vocoder suffers from a slight speech quality degradation, it still achieves an MOS of 4.5, which confirms the effectiveness of the WN vocoder for generating high-fidelity speech. For the collapse-free set, the results show that the vanilla WN vocoder (DMDN + WN) outperforms the WORLD vocoder (DMDN + WORLD), but the WN vocoder (DMDN + LPCDC), which always applies the LPCDC, suffers from a severe speech quality degradation. In addition, the results of the collapsed set indicate that the collapsed speech also significantly degrades the speech quality for both the DNN- and DMDN-based systems. The same MOSs of the collapse-free and collapsed sets generated by the DMDN-based system with the WN vocoder applying the LPCDC imply that although the LPCDC alleviates the collapsed speech problem, it causes extra speech quality degradation. However, applying the LPCDC with the CSSD, which limits the negative effect of the LPCDC to only the collapsed speech segments, not only markedly alleviates the collapsed speech problem but also prevents the WN vocoder from speech degradation caused by applying the LPCDC, so the system with LPCDC and CSSD in the collapsed set attains a similar MOS to the systems with the vanilla WN vocoder for the collapse-free set. In conclusion, the proposed LPCDC and CSSD modules significantly alleviate the collapsed speech problem of the WN vocoder while maintaining a similar speech quality.

### 2) SPEAKER SIMILARITY

Furthermore, we conducted a speaker similarity test on the proposed system, the DNN-based system with the vanilla WN vocoder, and the DMDN-based system with the WORLD vocoder. The listeners, devices, and environment were the same as in the MOS test. The same evaluation set including five subsets was adopted, and each subset was also evaluated by three listeners. The similarity measurement followed the speaker similarity test in VCC2018 [44], which asked listeners to listen to a natural target and a converted utterance and to determine whether the speakers of these two utterances are definitely the same, maybe the same, maybe different, or definitely different. The final similarity score is the sum of the definitely the same and maybe the same scores. As shown in Fig. 13, the proposed method achieves a higher speaker similarity than the DMDN-based system with the WORLD vocoder, which is consistent with previous comparisons with the WN and WORLD vocoders [37, 38]. Moreover, the proposed system also attains a similar speaker similarity to the DNN-based system with the WN vocoder, which confirms that the proposed LPCDC with the CSSD can simultaneously ease the collapsed speech problem, greatly alleviate the negative effect of the LPCDC, and maintain the same speaker

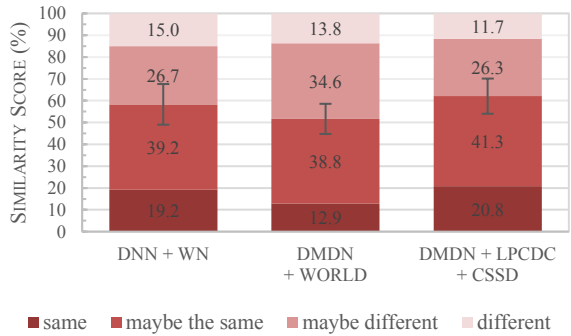

similarity as the vanilla WN vocoder without the collapsed speech problem.

### 3) COMPARISON WITH NU VCC2018 SYSTEM

Our non-parallel VC system submitted to VCC2018 (NU non-parallel VC system) [43] was a DNN-based non-parallel VC system with the vanilla WN vocoder. The collapsed speech utterances were detected using power differences between the generated and reference utterances, which achieves a lower accuracy than the CSSD method with waveform envelope detection. Moreover, the LPCDC was applied to the whole collapsed speech utterance, which also caused speech quality degradation. However, the system still attained second place in the speaker similarity test and above-average speech quality. In this paper, we utilize a DMDN-based model and propose the CSSD to apply the LPCDC to only the collapsed speech segments, which greatly alleviates the speech degradation caused by the LPCDC. The proposed system clearly outperforms the previously submitted system, and the experimental results also confirm the effectiveness of the proposed CSSD and LPCDC modules.

## VI. CONCLUSION

In this paper, we explored the phenomena, possible reasons, and negative effects of the collapsed speech problem of the WN vocoder. We also proposed the LPCDC technique to protect the WN vocoder from the collapsed speech problem, but it caused extra speech quality degradation. Therefore, we applied the CSSD to segmentally detect the collapsed speech and applied the LPCDC technique to only the detected segments, which greatly alleviated the speech degradation problem. To summarize, we proposed a system outperforming the previous system submitted to VCC2018.

## APPENDIX

For efficient WN generation, the waveform envelope and LPC coefficients of the WORLD-generated speech are extracted in advance, so only the WN vocoder sequentially generates the speech samples, which are segmentally checked by the CSSD in the testing stage. Furthermore, to simulate the effect of the μ–law codec, the WORLD-generated speech is also encoded and decoded by the μ–law.

The 8-bit μ–law encoding is as follows:

$$E(x) = \text{sgn}(x)\frac{\ln(1 + 255 * |x|)}{\ln(1 + 255)}, \qquad (10)$$

where $x$ is the input speech sample, and the output of the WN vocoder is the μ–law-encoded 256-level logistic distribution. Therefore, to create an LPC PMF, we first obtain the real waveform amplitude of each level $y_{level}$ as

$$y_{level} = E^{-1}(q), \qquad q \in [0, 1, ..., 255]. \ (11)$$

Then, the value of each level of the LPC PMF can be approximated as

$$lpc(y_{level}) = \frac{\exp(-((y_{level} - \mu_{lpc}) / \sigma_{lpc})^2 / 2)}{\sigma_{lpc}\sqrt{2\pi}}, \qquad (12)$$

where $\mu_{LPC}$ is the LPC-predicted value of the current sample and $\sigma_{LPC}^2$ is the variance of the LPC prediction errors. After we obtain the LPC PMF, we normalize it to make the summation equal to 1.